\documentclass[a4paper,fleqn]{cas-sc}

\usepackage[numbers,sort&compress]{natbib}
\usepackage{booktabs,multirow}
\usepackage{graphicx}
\usepackage{dcolumn}
\usepackage{bm}
\graphicspath{{./figures/}}
\usepackage[para,online,flushleft]{threeparttable}
\usepackage{soul}
\usepackage[normalem]{ulem}
\makeatletter
\expandafter\let\csname equation*\endcsname\relax
\expandafter\let\csname endequation*\endcsname\relax
\usepackage{amsmath}
\usepackage{tikz}
\usepackage{pgfplots}
\pgfplotsset{compat=1.18}
\usepgfplotslibrary{colorbrewer}
\usepgfplotslibrary{groupplots}

\usepackage{amsfonts,amssymb,amsthm}
\usepackage{xargs} 
\usepackage{bbold} 
\usepackage{scalerel} 

\usepackage{hyperref}
\usepackage{siunitx}
\DeclareSIUnit{\density}{kg/m^3}
\DeclareSIUnit{\stiffness}{N/m}
\DeclareSIUnit\radpersec{\radian/\second}
\usetikzlibrary{plotmarks, calc}
\usepackage{standalone}
\usepackage{subcaption}
\usepackage{physics}
\usepackage{float}
\usepackage{tabularx}
\usepackage{verbatim}
\usetikzlibrary{decorations}
\usetikzlibrary{decorations.markings}
\usetikzlibrary{plotmarks}
\usetikzlibrary{spy}
\restylefloat{table}		
\usepackage{pgfplots}
\pgfplotsset{compat=1.16}
\usepgfplotslibrary{groupplots}
\usepgfplotslibrary{colorbrewer}
\usetikzlibrary{patterns}
\usetikzlibrary{calc}
\usepackage{helvet} 
\usepackage{helvet} 
\usepgfplotslibrary{patchplots}
\newcommand{\bs}[1] {\bm{#1}}
\pgfplotsset{scaled y ticks=false}
\pgfplotsset{
colormap name=viridis,
}
\tikzset{
pattern size/.store in=\mcSize, 
pattern size = 5pt,
pattern thickness/.store in=\mcThickness, 
pattern thickness = 0.3pt,
pattern radius/.store in=\mcRadius, 
pattern radius = 1pt}
\makeatletter
\pgfutil@ifundefined{pgf@pattern@name@_9akz2l753}{
\pgfdeclarepatternformonly[\mcThickness,\mcSize]{_9akz2l753}
{\pgfqpoint{-\mcThickness}{-\mcThickness}}
{\pgfpoint{\mcSize}{\mcSize}}
{\pgfpoint{\mcSize}{\mcSize}}
{
\pgfsetcolor{\tikz@pattern@color}
\pgfsetlinewidth{\mcThickness}
\pgfpathmoveto{\pgfpointorigin}
\pgfpathlineto{\pgfpoint{0}{\mcSize}}
\pgfusepath{stroke}
}}
\makeatother
\tikzset{every picture/.style={line width=0.75pt}} 

\def\tsc#1{\csdef{#1}{\textsc{\lowercase{#1}}\xspace}}
\tsc{WGM}
\tsc{QE}
\tsc{EP}
\tsc{PMS}
\tsc{BEC}
\tsc{DE}

\begin{document}
\let\WriteBookmarks\relax
\def\floatpagepagefraction{1}
\def\textpagefraction{.001}

\shorttitle{Clamp-on systems' analysis}

\shortauthors{S V Valappil et~al.}

\title [mode = title]{A semi-analytical approach to characterize high-frequency three-dimensional wave propagation through clamp-on flowmeters} 



%
\author[1]{Sabiju Valiya Valappil}[type=author,
                        auid=000,bioid=1,
                        orcid=0000-0001-7511-2910]

\cormark[1]

\fnmark[1]

\ead{S.ValiyaValappil@tudelft.nl}
\ead[url]{https://www.tudelft.nl/en/2022/tnw/sabiju-valiya-valappil-joined-imphys-as-post-doc}
\credit{Conceptualization of this study, Methodology, Software, Writing - Original draft preparation}
\affiliation[1]{organization={Faculty of Applied Sciences, Delft University of Technology},
    addressline={Lorentzweg 1}, 
    city={Delft},
    postcode={2628 CJ}, 
    country={The Netherlands}}


\author%
[2]
{Alejandro M.~Arag\'{o}n}
\fnmark[2]
\ead{A.M.Aragon@tudelft.nl}
\ead[URL]{https://www.tudelft.nl/staff/a.m.aragon/}
\credit{Conceptualization, Supervision, Writing - Review \& Editing}
\author[2]{Johannes F. L. Goosen}[type=author]
\fnmark[3]
\ead{J.F.L.Goosen@tudelft.nl}
\ead[URL]{https://www.tudelft.nl/en/staff/j.f.l.goosen/}

\credit{Conceptualization, Supervision, Writing - Review \& Editing}
\affiliation[2]{organization={Faculty of Mechanical Engineering, Delft University of Technology},
    addressline={Mekelweg 2}, 
    city={Delft},
    postcode={2628 CD}, 
    country={The Netherlands}}

\cortext[cor1]{Corresponding author}



\begin{abstract}
Wave propagation analysis at high frequencies is essential for applications involving ultrasound waves, such as clamp-on ultrasonic flowmeters. However, it is extremely challenging to perform a 3D transient analysis of a clamp-on flowmeter using standard tools such as finite element analysis (FEA) due to the enormous associated computational cost. In this study, we separate the clamp-on flowmeter into different domains and analyze them separately. Wave propagation in the fluid domain is analyzed via FEA at low frequencies (\SI{100}{\kilo\hertz}, \SI{200}{\kilo\hertz}, and \SI{500}{\kilo\hertz}) and using ray tracing at high frequencies (\SI{1}{\mega\hertz}). The behavior in the solid domain (wedges and pipe wall) is analytically characterized via geometric projection. All these individual analyses provide us with different scaling factors with which the waves in the respective domains scale when 3D effects are considered. The complete clamp-on system is then analyzed in 2D via the Discontinuous Galerkin (DG) method to obtain the response at the receiver. The receiving signal is then scaled using the aforementioned scaling factors to accurately capture the wave propagation behavior of the clamp-on system in 3D. The output signal from the 2D analysis then becomes much clearer so that the fluid signal can be identified straightforwardly, which would be nearly impossible otherwise. 
\end{abstract}



\maketitle

\section{Introduction}
Ultrasound waves, because of their substantial power density, short wavelength, and non-invasive nature compared to electromagnetic waves, have been immensely applied in bio-imaging~\cite{Ibrahim_2013,988970,powers2011medical}, flowrate measurement~\cite{5007965,Moore_2000}, distance and velocity measurement~\cite{9817,BOLTRYK2004473}, temperature measurement~\cite{Motegi_2013}, and non-destructive evaluation~\cite{helal2015non,CHATILLON2000131}, among others. The global ultrasound device market size was valued at \$9.3 billion in 2022 and is expected to grow to \$15.4 billion by 2032~\cite{Allied_2022}. To effectively design ultrasound devices for different applications, it is necessary to characterize their behavior. However, experiments are generally expensive and time-consuming; thus, several analytical and numerical methods have been proposed.

Analytical solutions, although computationally efficient, exist only for simplified geometries such as 1D structures, cylinders, and spheres~\cite{achenbach2012wave,10.1121/1.415365,10.1121/1.415366,10.1063/1.1663783}. For more complicated geometries, numerical methods such as the finite element method (FEM)~\cite{8360151} or boundary element method (BEM)~\cite{TADEU2001499} are used. The former solves the governing equations after discretizing the domain in a finite number of interconnected elements of simple (usually linear) behavior, whereas the latter transforms the equations to integral form and solves them over the boundary. Although resulting in a small system size due to discretizing only the boundary, BEM's implementation is complex and has fully dense asymmetric matrices, in contrast to FEM's sparse banded symmetric matrices. FEM is generally used to solve wave propagation in solids and fluids because of its versatility in discretizing complex geometries~\cite{doi:10.1098/rspa.2016.0738}. However, the computational cost increases with an increase in frequency to account for the spatial and temporal accuracy (as a rule of thumb eight linear finite elements are required per wavelength)~\cite{rose2002baseline}. BEM and FEM have also been combined to leverage their respective advantages in analyzing wave propagation through a fluid-filled cylindrical shell, where a coupled vibration analysis is performed~\cite{ZHANG2001229}. 

The wave propagation analysis becomes particularly challenging when dealing with ultrasound devices such as clamp-on ultrasonic flowmeters operating at high frequencies (in \SI{}{\mega\hertz} range) and in far-field (where characteristic length is much higher than the acoustic wavelength). This is because the total number of degrees of freedom (DOFs) becomes enormous due to the large domain compared to the mesh size. Additionally, the time step required for the evaluation becomes very small at high frequencies (at least ten steps per period), leading to vast computational demands. In addition, in the case of time harmonic wave solutions, the accuracy of the numerical solution drastically decreases with the increase in wave number~\cite{IHLENBURG19959,Deraemaeker_1999}. For transient wave propagation, such as the case of the clamp-on flowmeter, the solution can exhibit spurious oscillations due to Gibb's phenomenon~\cite{doi:10.1137/S0036144596301390}, and dispersion and dissipation errors due to numerical period elongation and amplitude decay~\cite{https://doi.org/10.1002/eqe.4290010308}. When high-frequency waves travel long distances the cumulative error becomes large and the inaccuracy in the numerical solution rapidly increases~\cite{Idesman2011}. Additionally, the computed wave velocities may depend on the propagation direction due to the mesh, which causes waveform distortions, leading the model to behave anisotropically even in an isotropic medium.

Several approaches have been proposed to address the aforementioned challenges of FEM. In particular, the spectral method, which uses harmonic functions in the solution space, can obtain numerical solutions very close to exact solutions~\cite{gottlieb1977numerical}. However, the spectral method is difficult to implement for complex geometries since the method uses global basis functions. An extension of spectral method was proposed where high-order Lagrangian-based finite elements along with special integration scheme (Gauss-Lobatto-Legendre) were used. The method has low numerical dispersion compared to standard FEM~\cite{PATERA1984468} and is effective in explicit time integration. However, it still lacks means to model complex geometries, which was addressed in parts by combining spectral method with FEM---spectral finite element method~\cite{Gopalakrishnan2011,10.1115/1.2203338}.     

A second method to overcome challenges of FEM is by means of higher-order Discontinuous Galerkin (DG) method, which lead to block-diagonal mass matrices~\cite{hesthaven2007nodal,doi:10.1137/S1064827598343723}. The higher order is essential here since the wave equations produce parasitic waves, which should be suppressed by using penalty terms that are dissipative. Thus, DG possess low numerical dissipation and dispersion~\cite{WILCOX20109373}. Moreover, DG is also capable of avoiding the Runge phenomenon for higher-order polynomial approximations~\cite{doi:10.1137/S003614299630587X}. DG also facilitates non-conformal discretizations allowing a flexible meshing schemes for fluid and solid domains~\cite{doi:10.1137/05063194X}. In other words, due to the discontinuous interpolation space and the use of higher-order shape functions (4\textsuperscript{th} to 6\textsuperscript{th} polynomial order), we can have non-matching interfaces and use a significantly smaller number of elements (as a rule of thumb 1.5 to 2 elements per wavelength). DG has been used in linear and non-linear ultrasound simulations including elastic and acoustic waves~\cite{10.1093/imanum/drab089,Chou2017,10.1121/1.5032196,8092235}.
However, even with DG the 3D transient analysis of a clamp-on ultrasonic flowmeter is still computationally very intensive due to the different types of wave interactions between large-sized domains. To the best of our knowledge, there are no known methods that can characterize elastic and acoustic wave propagation through a large system such as a clamp-on flowmeter in 3D at high frequencies.


In this study, we propose an approach that combines DG, ray tracing, and geometric projection methods to provide an accurate interpretation of the complete wave propagation through a clamp-on flowmeter. We perform 3D DG analysis only for the fluid domain at low frequencies (\SI{100}{\kilo\hertz}, \SI{200}{\kilo\hertz}, and \SI{500}{\kilo\hertz}). The behavior at high frequencies is obtained qualitatively via 3D ray tracing, which provides a scaling factor. The spread of the acoustic field on the pipe wall is then evaluated by the geometric projection. Combining these three responses with a 2D model of the complete clamp-on system provides us with its full wave propagation behavior.   

\section{Problem definition}
\label{sec:problem}
In this section, we discuss the different types of waves present in a clamp-on ultrasonic flowmeter, the wave interactions, and the resulting wavefields. The properties and operation of the clamp-on ultrasonic flowmeter are discussed in detail in our previous publication~\cite{VALAPPIL2025112173}. We take the geometric and material properties from there as listed in Table~\ref{tab:clamp_on_properties}.
\begin{table}[!ht]
    \centering
    \begin{tabular}{|c|c|c|c|c|c|}\hline
        \multicolumn{6}{|c|}{\textbf{Pipe (stainless steel) properties}}\\
        \hline
        $D_o=\SI{80}{\milli\meter}$ & $t_p=\SI{4}{\milli\meter}$ & $L_p=\SI{140}{\milli\meter}$ & $\rho_p=\SI{7800}{\density}$ & $c_{pp}=\SI[per-mode=symbol]{4935.5}{\meter\per\second}$ & $c_{ps}=\SI[per-mode=symbol]{3102.9}{\meter\per\second}$
        \\ \hline
        \multicolumn{6}{|c|}{{\textbf{Wedge (polysulfone) properties}}}\\
        \hline
        $b_w=\SI{36}{\milli\meter}$ & $h_w=\SI{22.5}{\milli\meter}$ & $t_w=\SI{30}{\milli\meter}$ & $\rho_w=\SI{1350}{\density}$ & $c_{wp}=\SI[per-mode=symbol]{1958.84}{\meter\per\second}$ & $c_{ws}=\SI[per-mode=symbol]{1183.38}{\meter\per\second}$\\
        \hline
    \end{tabular}
    \caption{Geometric and material properties of the pipe and wedge, where $D_o$, $t_p$, and $L_p$ respectively, are the outer diameter, the wall thickness, and the length of the pipe. $b_w$, $h_w$, and $t_w$ are the base, height, and thickness of the wedge, respectively. $\rho_p$, $c_{pp}$, and $c_{ps}$, respectively, are the density, pressure, and shear wave speeds of the pipe, while $\rho_w$, $c_{wp}$, and $c_{ws}$ are the corresponding material properties of the wedge.}
    \label{tab:clamp_on_properties}
\end{table}
Polysulfone was selected as the wedge material, stainless steel for the pipe, and water (density, $\rho_f=\SI{998.2}{\density}$ and sound speed, $c_f=\SI[per-mode=symbol]{1481.4}{\meter\per\second}$) for the fluid.
\begin{figure}[!ht]
    \centering
    \def\svgwidth{1\linewidth}
	\input{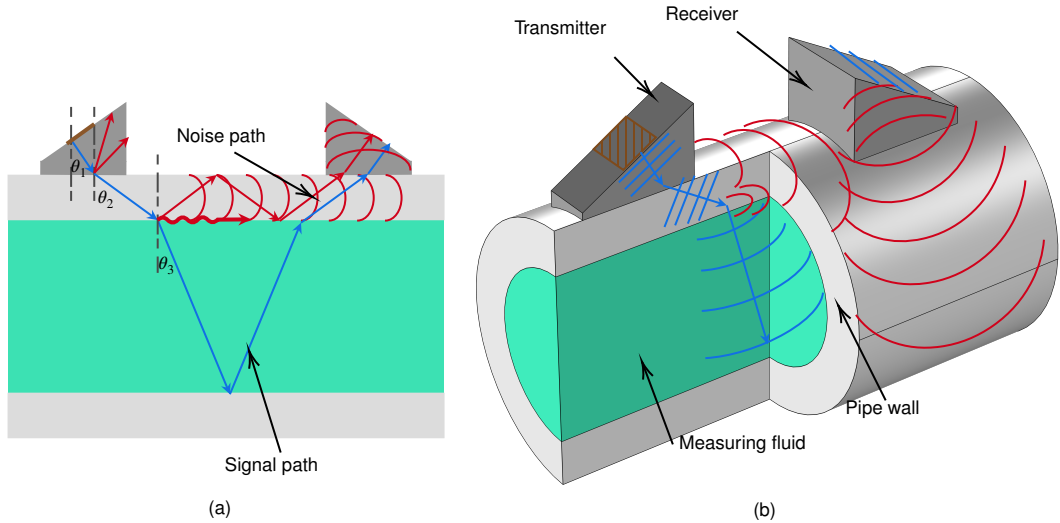}
    \caption{Schematic representation of the wave propagation through a clamp-on ultrasonic flowmeter (a) 2D view with signal (blue) and noise (red) paths along with different refraction angles. The incoming angle $\theta_1=\SI{30}{\degree}$, refraction angle at the wedge-pipe wall interface $\theta_2=\SI{52.4}{\degree}$, and transmission angle from the pipe to fluid $\theta_3=\SI{22.2}{\degree}$ are also marked here (b) The 3D representation of the wavefield from (a), where different components of the clamp-on ultrasonic flowmeter are labeled. The brown hatched region is the location of the input signal (prescribed displacement). Schematics of 3D representations of the wavefields in the fluid, solid, and interface are also provided here.}
    \label{fig:schematic}
\end{figure}
Figure~\ref{fig:schematic} shows a schematic representation of the different wave paths and the possible wavefields of a fluid-filled cylinder excited by an oblique source of ultrasound pulse. 

In Figure~\ref{fig:schematic}(a), the blue arrows show the required signal path, while the red arrows and curves represent the noise generated along different interfaces. The 2D wave propagation through the clamp-on system is also discussed in detail in the same article. Here, we emphasize the aspects that the 2D analysis cannot comprehend, for instance, due to its intrinsic restriction, the 2D model cannot predict the radial expansion of the wavefield in the fluid domain as represented in Figure~\ref{fig:schematic}(b). The figure also shows a more detailed representation of signal and noise paths.

The input signal in the transmitter is assumed to be a plane wave, which is a fair approximation as the pulse is coming from a piezoelectric crystal attached to the transmitting wedge (see the brown hatched region in the figure). The incoming plane wave experiences several reflections and refractions generating multiple signals before reaching the receiver. Due to the cylindrical shape of the pipe and fluid, the primary signal (blue) entering the fluid domain expands radially before impinging the opposite pipe wall. This signal then converges on a region within the fluid domain before arriving at the receiver. Similarly, the primary noise signal traveling via the pipe wall also spreads around the wall before reaching the receiver as shown in Figure~\ref{fig:schematic}(b). Additionally, the ringing down of waves among the wedges, and interference between different noise pulses and the required signal also experience the influence of the 3\textsuperscript{rd} dimension and will scale accordingly. Thus, we explore different methods to analyze the 3D wave propagation through the clamp-on system as discussed next.
\section{Methods for high-frequency wave propagation analysis in a clamp-on flowmeter}
As the high-frequency analysis of the clamp-on flowmeter is incredibly complex, we segment the geometry into different domains and analyze them separately. To that end, we select multiple methods, investigate their pros and cons, and possibilities to combine their responses as below.
\subsection{Geometric acoustics}
Ray tracing is a geometric acoustic method that can compute trajectories, phase, and intensity of acoustic rays, which is valid in the high-frequency limit where the acoustic wavelength is much smaller than the characteristic length of the geometry~\cite{Pierce2019_8}. Using the trajectory, we can obtain the arrival time and location of the acoustic pulse at the receiver. We can also calculate the scattering and absorption of the traveling waves using ray tracing. Additionally, it provides the focal location within the medium with a curved boundary (such as the current situation). Nonetheless, ray tracing assumes the waves to be straight lines traveling without any distortions, resulting in variations when dealing with complex geometries, dispersive environments, and wave interference. Thus we can apply ray tracing to the fluid domain, which has a more uniform geometry and acoustic response compared to the solid domain. The equation that governs the ray acoustics takes the following form:
\begin{align}
    \frac{d\bm{k}}{dt}=\frac{\partial \omega}{\partial \bm{x}} && \frac{d\bm{x}}{dt}=-\frac{\partial \omega}{\partial \bm{k}},
    \label{eq:ray_tracing}
\end{align}
where $\bm{k}$ is the wave vector, $\omega$ is the circular frequency (in \SI[per-mode=symbol]{}{\radian\per\second}), $\bm{x}$ is the spatial coordinate vector, and $t$ is the time coordinate. The boundary condition (BC) necessary to solve the system of equations is provided by a prescribed wave vector as follows:
\begin{equation}
    \bm{k}(\bm{x}_0)=k_0\frac{\bm{L}_0}{|\bm{L}_0|},
    \label{eq:ray_BC}
\end{equation}
where $k_0$ is the magnitude of the wave vector prescribed at $\bm{x}_0$ and $\frac{\bm{L}_0}{|\bm{L}_0|}$ is the unit ray direction vector. Using the ray equation~\eqref{eq:ray_tracing} and BC~\eqref{eq:ray_BC} we can set up the analysis to obtain the wave propagation characteristics in the fluid domain.

Within the solid domain, the wedge mostly has straight boundaries except for the interface with the pipe wall, which in most cases can also be assumed straight due to the significant dimensional difference between the two (the width and thickness of the wedge are considerably smaller than the diameter of the pipe). Hence the primary pulse traveling through the wedge accurately follows Snell's law and can be characterized analytically. The secondary reflections and refractions at the wedge-pipe wall interface can be precisely analyzed via 2D FEA as discussed in Section~2.3 of ~\cite{VALAPPIL2025112173}. Thus, we do not need additional methods to characterize the wedge. On the contrary, the pipe wall allows radial expansion of the acoustic waves, and hence by projecting the expanding wave field from a cylinder to a plane, we determine the proportion of energy reaching the receiver. The assumptions here are that the pipe wall is non-dispersive and non-dissipative, which is fairly valid considering the pipe is made of stainless steel. The proportion factor is then used to scale the acoustic signal from the primary noise extracted at the receiver using the 2D FEA. To obtain a better understanding of the traveling waves in the clamp-on system, we also analyze it using 3D FEA as discussed next.
\subsection{3D finite element analysis}
\begin{figure}[!ht]
    \centering
    \def\svgwidth{0.75\linewidth}
	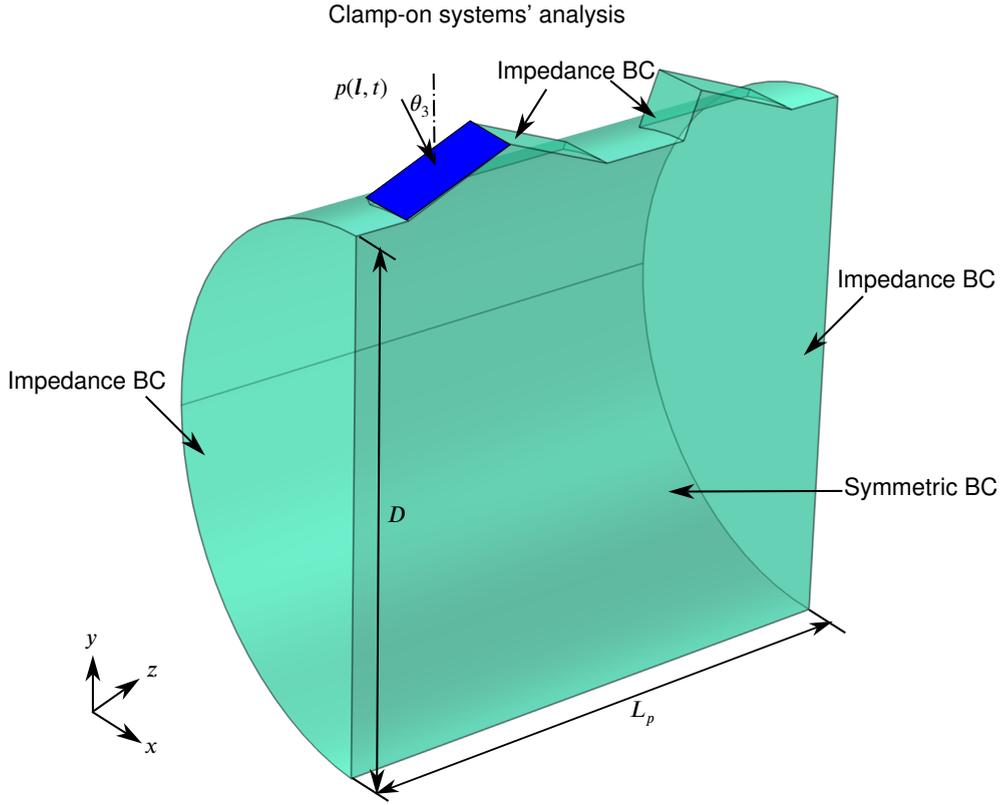
    \caption{3D model of the fluid domain with necessary BCs. $p(\bm{l},t)$ is the prescribed pressure applied at an angle $\theta_3$ while impedance and symmetric BCs are provided on the flat surfaces of the fluid domain except for transmitter and receiver locations. The inner diameter of the pipe $D=D_o-2\times t_p=\SI{72}{\milli\meter}$ and length $L_p=\SI{140}{\milli\meter}$ are also marked here.}
    \label{fig:3D_model}
\end{figure}

As the 3D FEA of the complete clamp-on system at \SI{1}{\mega\hertz} is computationally demanding even within a high-performance computing cluster, we reduce the complexities by removing the solid domain. Additionally, by performing the acoustic analysis at low frequencies (\SI{100}{\kilo\hertz}, \SI{200}{\kilo\hertz}, and \SI{500}{\kilo\hertz}), we can obtain the 3D acoustic response within the fluid medium. Figure~\ref{fig:3D_model} shows the geometry of the fluid domain with the necessary BCs used to perform the 3D FEA.
Since the domain is symmetric to the $y-z$ plane, we only need to analyze half of the domain allowing us to reduce the computational cost further. Thus the design shown in the figure is a semi-circular cylinder with two wedges on top to mimic the transmitter and receiver. As the incoming pressure wave enters the fluid from the pipe wall at an angle $\theta_3$ (refer Figure~\ref{fig:schematic}(a)), we modify the geometry to accommodate for the same (both transmitter and receiver). 
We perform a time-domain analysis using DG FEM with Comsol to characterize the wave propagation in the fluid domain for which the linearized Euler equations are solved. The acoustic wave equation in the absence of body forces takes the form:
\begin{equation}
    \begin{aligned}
       \Delta p_t=\rho_f \kappa \Ddot{p_t}, && \Delta \bm{v}_f=\rho_f \kappa \Ddot{\bm{v}_f},
    \label{eq:acoustic_wave}
    \end{aligned}
\end{equation}
where $p_t$ is the total acoustic pressure, $\bm{v}_f$ is the total acoustic velocity, $\kappa$ is the compressibility, and $\Delta$ is the Laplacian operator in 3D. We still need to supply the necessary BCs to solve this boundary value problem. We provide a Dirichlet BC (prescribed pressure) along the boundary of the fluid domain shaded using a blue quadrilateral in Figure~\ref{fig:3D_model}, which takes the form
\begin{equation}
    p_t(\bm{l},t)= \bar{p}g(t), \hspace{3mm} \text{applied at} \hspace{3mm} \underline{x} = \underline{l},
    \label{eq:Dirichlet}
\end{equation}
where, $\bar{p}$ is the constant pressure amplitude that is multiplied by $g(t)$, which is a broadband Gaussian pulse with a \SI{100}{\percent} \SI{-6}{\decibel} bandwidth.
The two flat surfaces of the fluid domain normal to the $z$ axis are supplied with impedance BCs, which mimic a low-reflective boundary allowing the waves to escape through those boundaries. The impedance BC takes the form
\begin{equation}
    \bm{n}\cdot\bm{v}_f=\frac{p_t}{Z_i},
    \label{eq:impedance}
\end{equation}
where $\bm{n}$ is the unit outward normal vector on the surface and $Z_i=\rho_f c_f$ is the acoustic impedance of the fluid medium. We also provide impedance BCs to the flat surfaces of the wedges except for the input (blue-shaded plane) and output (similar plane on the receiver) surfaces to limit the reflections. We further apply a symmetric BC on the flat surface along the $z$ axis, which takes the form
\begin{equation}
    \bm{n}\cdot\bm{v}_f=0.
    \label{eq:symmetric}
\end{equation}
The curved surface of the water domain and the flat surface of the receiver that is at $\theta_3$ to $z$ axis are left free allowing the waves a complete reflection from those boundaries. The details of these BCs are available in Comsol documentation~\cite{comsol6}. Using wave equation~\eqref{eq:acoustic_wave} and BCs \eqref{eq:Dirichlet}, \eqref{eq:impedance}, and \eqref{eq:symmetric}, we can set up the numerical analysis to obtain the wave propagation behavior of the fluid domain. These responses can then be compared against the ray acoustics and 2D FEA results to determine the 3D scaling factor. 

However, to apply all the aforementioned scaling factors effectively, we need to know the arrival times of each of these signals in the receiver. Arrival times of the required signal, primary noise, and ring-down noise can be determined by their travel distance and sound speeds in different materials. The arrival times of noise produced by the interference of different waves within the wall and the wall-fluid interface are more challenging and can only be obtained accurately via a 3D analysis of the entire geometry. This is out of the scope of this study and hence omitted from further investigation. 
By performing analytical and numerical analyses on the clamp-on geometry, we calculate various responses of the clamp-on system as discussed next.
\section{Results}
\subsection{Ray tracing response of the fluid medium}
\label{subsec:ray_tracing}
\begin{figure}[!ht]
    \centering
    \def\svgwidth{0.9\linewidth}
	\input{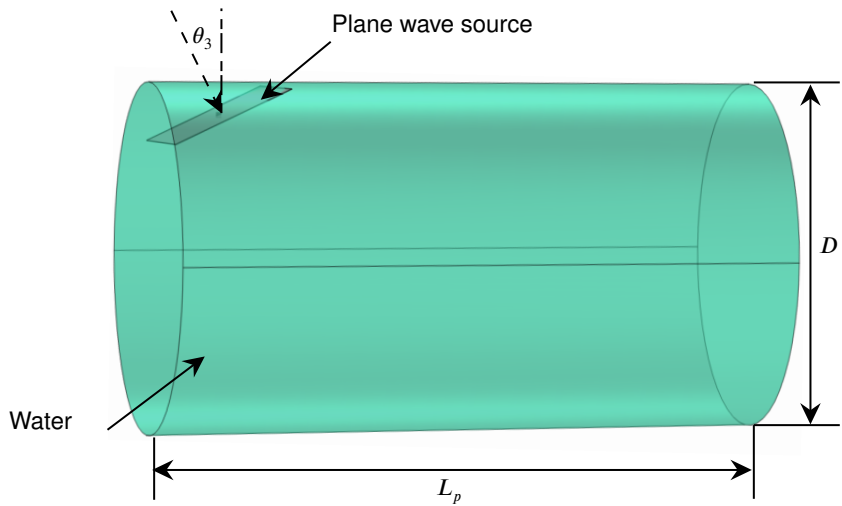}
    \caption{Water domain geometry used in the ray tracing analysis where the inclined plane shows the wave source at an angle $\theta_3$ to the normal. The diameter $D$ and length $L_p$ of the fluid domain are labeled here.}
    \label{fig:ray_tracing}
\end{figure}

Ray tracing analysis is performed using Comsol multiphysics' ray acoustics module, for the geometry shown in Figure~\ref{fig:ray_tracing}. Here, we assume no attenuation of the waves within the domain and that the boundaries have specular reflections, i.e., the waves impinging the boundaries reflect at the same angle~\cite{kuttruff2016room}.
The acoustic source selected here is a plane wave oriented at $\theta_3=\SI{22.2}{\degree}$ to the normal, which is the transmission angle of the pressure pulse across pipe wall-fluid interface (as described in Figure~\ref{fig:schematic}(a)). The diameter $D=\SI{72}{\milli\meter}$ and the length of the fluid domain $L_p$ (same as the pipe) are marked in the figure. Since at \SI{1}{\mega\hertz} the acoustic wavelength in water $\lambda=c_w/f=\SI{1.5}{\milli\meter}$ is much smaller than $D$, ray acoustics is valid for this situation. We perform the analysis for $\SI{120}{\micro\second}$ with a time step of \SI{0.2}{\micro\second} and the resulting ray trajectories are studied. 

\begin{figure}[!ht]
    \centering
    \def\svgwidth{1\linewidth}
	\input{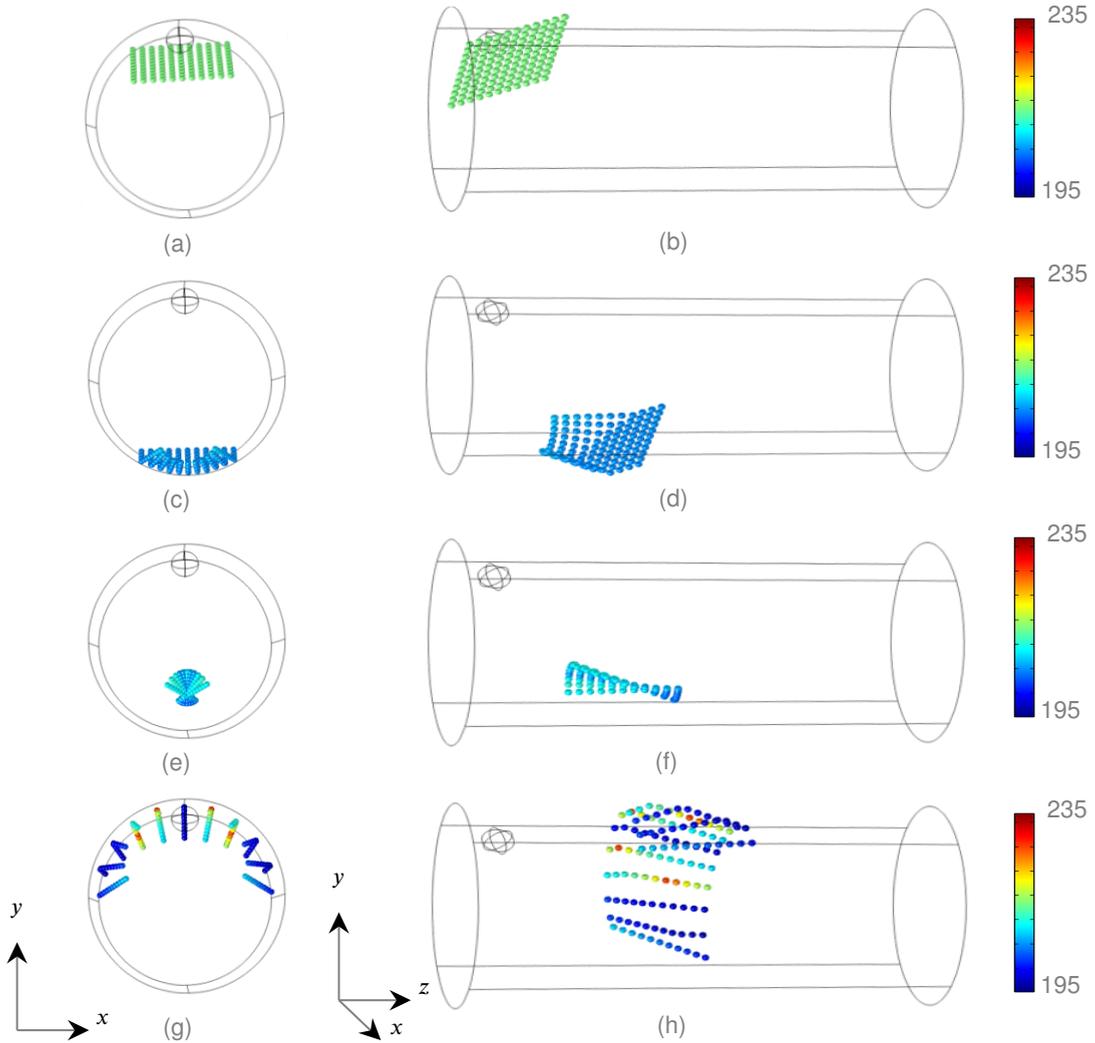}
    \caption{Ray trajectories through the water domain with an oblique plane source. The cross-sectional views of the domain correspond to (a) the start of the simulation, (c) when the rays impinge the opposite wall and start to reflect, (e) the rays focus within the medium, and (g) when the rays reach the upper boundary (close to the receiver). (b), (d), (f), and (h), respectively, are the corresponding longitudinal-sectional views. The color of the dots indicates the sound pressure level in \SI{}{\decibel} at that instance.}
    \label{fig:ray_tracing_res}
\end{figure}

Figure~\ref{fig:ray_tracing_res} shows different instances of the ray trajectories where we can clearly see the focusing and diverging effects of the waves within the cylindrical domain. The source is described as a grid with 121 points (11 points per side of a square), each of which travels at the same speed within the domain. The color of the dots represents the sound pressure level (SPL) in \SI{}{\decibel}.  
When the rays are traveling from the source to the opposite wall, they have the same SPL and are within the same plane as shown in Figures~\ref{fig:ray_tracing_res}(a) to~\ref{fig:ray_tracing_res}(d). However, they impinge the curved wall at different times as shown in Figure~\ref{fig:ray_tracing_res}(d) where some points have touched the wall while others have not. This results in a change in their wavefront shape from a plane to concave (see Figure~\ref{fig:ray_tracing_res}(f)). Since a curved wavefront keeps on expanding/contracting to maintain the energy, this concave wavefront shrinks to focus on a small region as represented in Figure~\ref{fig:ray_tracing_res}(e). After focusing, the wavefront changes its shape to convex and expands during its travel to the wall as can be seen in Figures~\ref{fig:ray_tracing_res}(g) and \ref{fig:ray_tracing_res}(h). Noteworthy, due to the focusing effect, some dots within the wavefront possess higher SPL compared to the rest as evident from the same figures. Additionally, we can also see that after the expansion some parts of the wavefront impinge on the upper wall and travel back to the fluid domain. These rays can interact with the outer wall and also interfere among themselves creating more noise. However, since they have lower SPL and arrive after the required signal, they can be identified and filtered out from the output signal. 

\begin{figure}[!ht]
    \centering
    \def\svgwidth{0.9\linewidth}
\begingroup%
  \makeatletter%
  \providecommand\color[2][]{%
    \errmessage{(Inkscape) Color is used for the text in Inkscape, but the package 'color.sty' is not loaded}%
    \renewcommand\color[2][]{}%
  }%
  \providecommand\transparent[1]{%
    \errmessage{(Inkscape) Transparency is used (non-zero) for the text in Inkscape, but the package 'transparent.sty' is not loaded}%
    \renewcommand\transparent[1]{}%
  }%
  \providecommand\rotatebox[2]{#2}%
  \newcommand*\fsize{\dimexpr\f@size pt\relax}%
  \newcommand*\lineheight[1]{\fontsize{\fsize}{#1\fsize}\selectfont}%
  \ifx\svgwidth\undefined%
    \setlength{\unitlength}{1248.88278751bp}%
    \ifx\svgscale\undefined%
      \relax%
    \else%
      \setlength{\unitlength}{\unitlength * \real{\svgscale}}%
    \fi%
  \else%
    \setlength{\unitlength}{\svgwidth}%
  \fi%
  \global\let\svgwidth\undefined%
  \global\let\svgscale\undefined%
  \makeatother%
  \begin{picture}(1,0.26084708)%
    \lineheight{1}%
    \setlength\tabcolsep{0pt}%
    \put(0,0){\includegraphics[width=\unitlength,page=1]{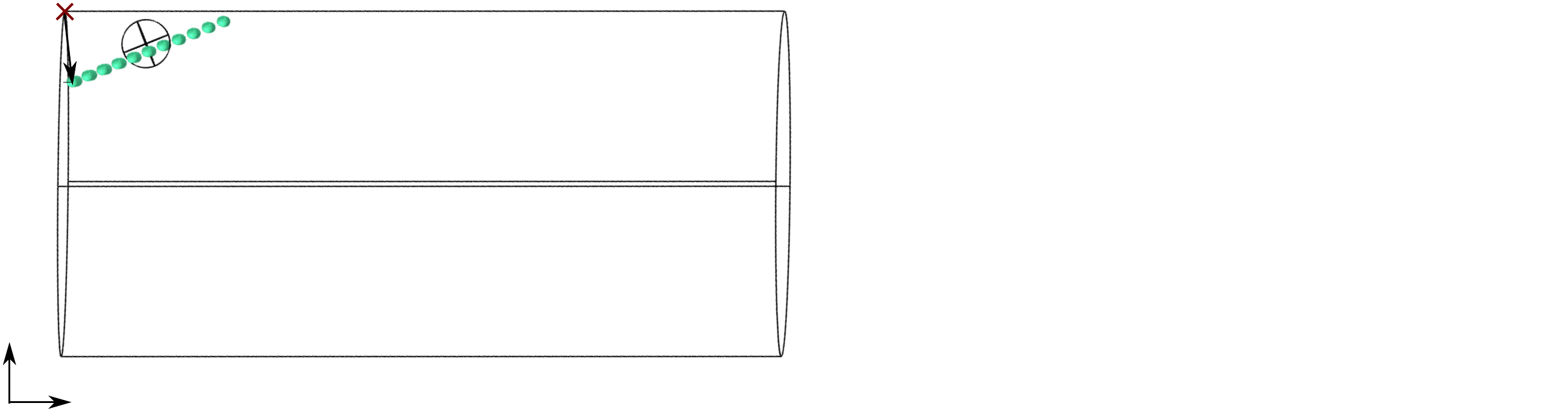}}%
    \put(0.04704175,0.00351317){\makebox(0,0)[lt]{\lineheight{5}\smash{\begin{tabular}[t]{l}$z$\end{tabular}}}}%
    \put(-0.00066466,0.04795087){\makebox(0,0)[lt]{\lineheight{5}\smash{\begin{tabular}[t]{l}$y$\end{tabular}}}}%
    \put(0.2568056,0.00566682){\makebox(0,0)[lt]{\lineheight{5}\smash{\begin{tabular}[t]{l}(a)\end{tabular}}}}%
    \put(0.76102341,0.00559482){\makebox(0,0)[lt]{\lineheight{5}\smash{\begin{tabular}[t]{l}(b)\end{tabular}}}}%
    \put(0,0){\includegraphics[width=\unitlength,page=2]{figs/ray_tracing_focal.pdf}}%
  \end{picture}%
\endgroup%

    \caption{Side views of ray trajectories for (a) the start of the simulation and  (b) when the rays are focusing in the fluid medium. The maroon cross is the reference point in both (a) and (b), while the arrows indicate the distance from the reference to the beginning of (a) source and (b) focal region.}
    \label{fig:ray_focal}
\end{figure}

From the ray trajectory, we can estimate the approximate location of the focal region with respect to the source as described in Figure~\ref{fig:ray_focal}.
Here two instances of the ray projection are shown; Figure~\ref{fig:ray_focal}(a) the start of the simulation when the rays have not started propagating yet, and Figure~\ref{fig:ray_focal}(b) when the rays are focusing after reflecting from the bottom wall. The top left corner of the pipe wall is selected as the reference coordinate, which is represented using a maroon cross in both figures. For simplicity, we track the movement of the first row of dots (sources) to estimate the focal region. The coordinates of the selected source points in the $y, z$ plane measured in \SI{}{\milli\meter} are (1.81, 15.04), while the coordinates of the focal point are (32.66, 54.62). Thus the distance between the source and the focal point for the first raw of dots is \SI{50.2}{\milli\meter}. Noteworthy, with respect to the source's location and orientation, the position of the focal region changes. However, since the orientation of the source is consistent (kept to $\theta_3$), the relative position between the source and the focal region follows this prediction. The focal length information can be used as a design parameter to optimize the clamp-on flowmeter's performance since by shifting the focal point we can change the arrival signal strength.

\begin{figure}[!ht]
    \centering
    \def\svgwidth{0.8\linewidth}
\begingroup%
  \makeatletter%
  \providecommand\color[2][]{%
    \errmessage{(Inkscape) Color is used for the text in Inkscape, but the package 'color.sty' is not loaded}%
    \renewcommand\color[2][]{}%
  }%
  \providecommand\transparent[1]{%
    \errmessage{(Inkscape) Transparency is used (non-zero) for the text in Inkscape, but the package 'transparent.sty' is not loaded}%
    \renewcommand\transparent[1]{}%
  }%
  \providecommand\rotatebox[2]{#2}%
  \newcommand*\fsize{\dimexpr\f@size pt\relax}%
  \newcommand*\lineheight[1]{\fontsize{\fsize}{#1\fsize}\selectfont}%
  \ifx\svgwidth\undefined%
    \setlength{\unitlength}{366.33054718bp}%
    \ifx\svgscale\undefined%
      \relax%
    \else%
      \setlength{\unitlength}{\unitlength * \real{\svgscale}}%
    \fi%
  \else%
    \setlength{\unitlength}{\svgwidth}%
  \fi%
  \global\let\svgwidth\undefined%
  \global\let\svgscale\undefined%
  \makeatother%
  \begin{picture}(1,0.59585318)%
    \lineheight{1}%
    \setlength\tabcolsep{0pt}%
    \put(0,0){\includegraphics[width=\unitlength,page=1]{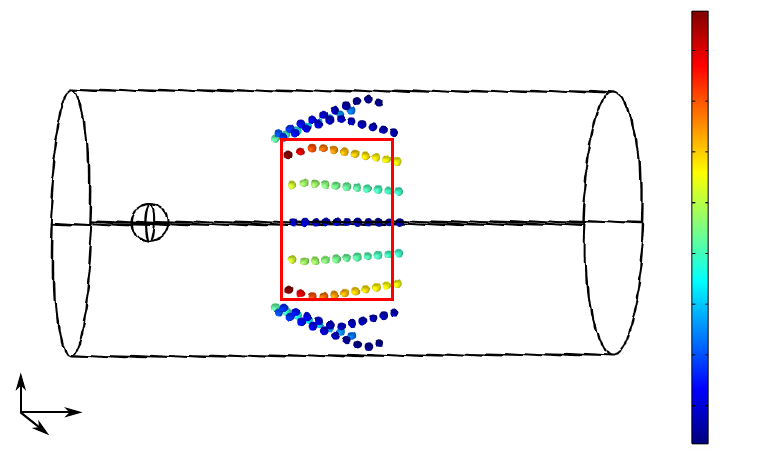}}%
    \put(0.10912245,0.04739457){\makebox(0,0)[lt]{\lineheight{5}\smash{\begin{tabular}[t]{l}$z$\end{tabular}}}}%
    \put(0.0217338,0.11728072){\makebox(0,0)[lt]{\lineheight{5}\smash{\begin{tabular}[t]{l}$x$\end{tabular}}}}%
    \put(0.06638505,0.0097728){\makebox(0,0)[lt]{\lineheight{5}\smash{\begin{tabular}[t]{l}$y$\end{tabular}}}}%
    \put(0.93572015,0.01261236){\makebox(0,0)[lt]{\lineheight{5}\smash{\begin{tabular}[t]{l}196\end{tabular}}}}%
    \put(0.93921873,0.56534227){\makebox(0,0)[lt]{\lineheight{5}\smash{\begin{tabular}[t]{l}214\end{tabular}}}}%
  \end{picture}%
\endgroup%

    \caption{Top view of the ray trajectory when the rays impinge the top wall where the red rectangle represents the area of the receiver. The color chart captures the SPL of each ray.}
    \label{fig:ray_output_signal}
\end{figure}
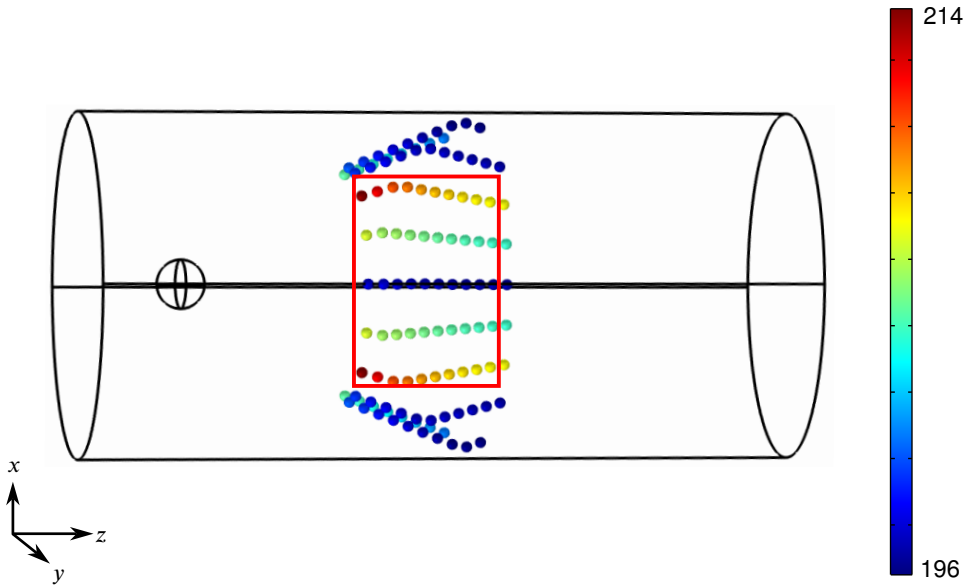

Using this ray acoustics model, we can also qualitatively estimate the energy of the signal at the receiver by counting the number of dots that reached the receiver and convoluting them with the corresponding SPLs. For instance, here approximately 50 out of 121 dots have reached the receiver location as highlighted by the red box in Figure~\ref{fig:ray_output_signal}, which shows the top view of the ray trajectory.
From the number of dots alone, we can see that the total energy decreases by a factor $f_1=2.4$. Their SPLs are in the range of \SI{196}{\decibel} to \SI{214}{\decibel} (refer to the color bar from the same figure). Calculating the sum of each dot's SPL values and subtracting from the average SPL of the center raw provides the change in SPL within the enclosure to be $\Delta$SPL=\SI{7.7}{\decibel}. This factor can be incorporated into the aforementioned scaling factor $f_1$ to obtain the influence of radial expansion, focusing, and divergence of the acoustic field within the fluid domain. To that end, we convert the relative SPL value to the fluid pressure ratio $p_f$ using the following expression
\begin{align}
    \text{SPL} = 20\log p_f, && p_f=10^{\left(\text{SPL}/20\right)},
    \label{eq:SPL}
\end{align}
and we get $p_f=2.42$. By dividing $p_f$ by $f_1$, we obtain an approximate scaling factor for the fluid domain, $p_r=1.01$. In other words, the output signal at the receiving location should be scaled in magnitude by the factor $p_r$ if the effects of 3D are not considered in the analysis. We further analyze the fluid domain through 3D FEA to compare the responses between ray tracing and FEA.

\section{3D finite element analysis of the fluid domain}
3D FEA in the time domain is performed on the geometry shown in Figure~\ref{fig:3D_model} for three different central frequencies, $\SI{100}{\kilo\hertz}$, $\SI{200}{\kilo\hertz}$, and $\SI{500}{\kilo\hertz}$ frequencies. A time step of $T/10$, where $T$ is the time period corresponding to the center frequency of the incoming pulse is used throughout the analysis. Since it takes about \SI{120}{\micro\second} for the pulse to travel from the transmitter to the receiver in a non-dispersive situation (sound speed independent of the frequency), we select the total simulation time for all three cases to be \SI{125}{\micro\second}. We compare the pressure profile, arrival time, amplitude, and signal spread (in time) between these analyses to obtain the acoustic behavior of the fluid domain in low and medium frequencies.
\subsection{Pressure profile comparison between different frequencies}
\label{subsec:pressure_comparison_3D}
\begin{figure}[!ht]
    \centering
    \def\svgwidth{1\linewidth}
	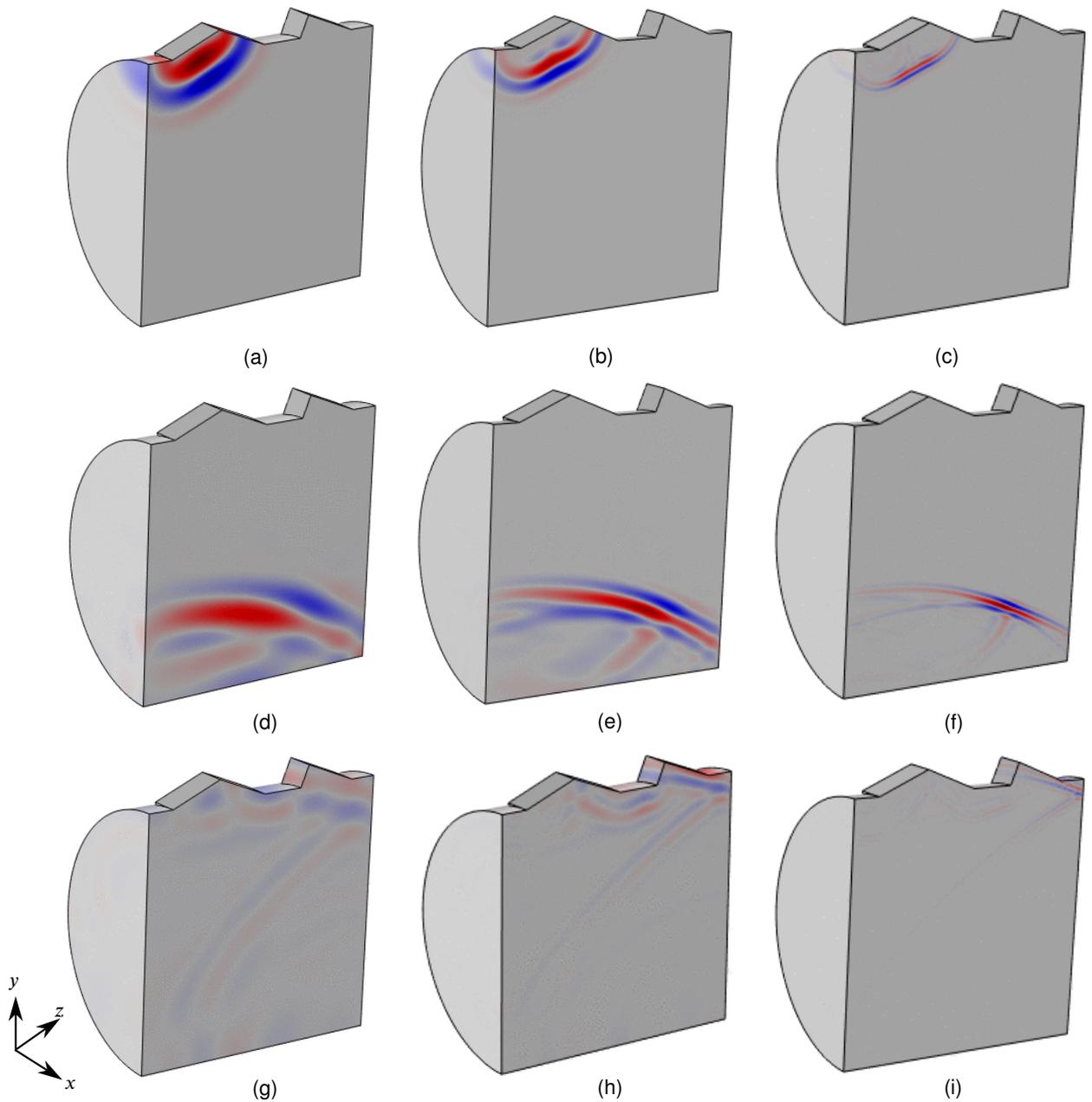
    \caption{Pressure profiles at different instances for 3D wave propagation in water at different frequencies. (a), (d), and (g), respectively, represent the start of the wave propagation, when the acoustic waves focus within the domain, and when it reaches the receiver at \SI{100}{\kilo\hertz}. (b), (e), and (h) are the corresponding behavior at \SI{200}{\kilo\hertz}, whereas (c), (f), and (i), respectively, are the behavior at \SI{500}{\kilo\hertz}.}
    \label{fig:pressure_profiles}
\end{figure}
Figure~\ref{fig:pressure_profiles} shows the comparison of the total pressure of the fluid domain at different instances corresponding to \SI{100}{\kilo\hertz}, \SI{200}{\kilo\hertz}, and \SI{500}{\kilo\hertz}. The instances shown here include the starting of the simulation (Figures~\ref{fig:pressure_profiles}(a), \ref{fig:pressure_profiles}(b), and \ref{fig:pressure_profiles}(c)), when the waves focus within the domain (Figures~\ref{fig:pressure_profiles}(d), \ref{fig:pressure_profiles}(e), and \ref{fig:pressure_profiles}(f)), and when the pulse arrives at the receiver (Figures~\ref{fig:pressure_profiles}(g), \ref{fig:pressure_profiles}(h), and \ref{fig:pressure_profiles}(i)). As we know, a low-frequency pulse experiences a larger spread due to diffraction~\cite{Pierce2019_9}, we see that the signal spread is higher for \SI{100}{\kilo\hertz} from the beginning of the simulation as compared to the other two cases (see Figures~\ref{fig:pressure_profiles}(a), \ref{fig:pressure_profiles}(b), and \ref{fig:pressure_profiles}(c)). The \SI{200}{\kilo\hertz} case although spreads relatively less, still focuses on a larger region (\ref{fig:pressure_profiles}(e)), leading to loss of energy when reaching the receiver. On the contrary, the \SI{500}{\kilo\hertz} signal experiences little spread while traveling through the fluid and focuses on a smaller region and reaches the receiver as shown in Figures~\ref{fig:pressure_profiles}(c), \ref{fig:pressure_profiles}(f), and \ref{fig:pressure_profiles}(i). This is also the reason why high frequency signals are preferred during the measurement process. To further investigate the signal response, we measure the output pulse at the receiving location of the fluid domain and compare them, as follows.

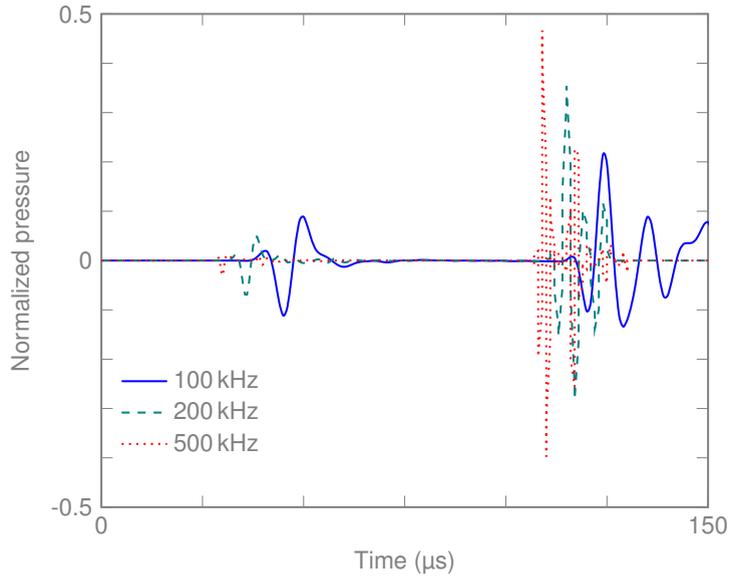
\begin{figure}[!ht]
    \centering
    \def\svgwidth{1\linewidth}
	\pgfplotsset{cycle list/Dark2}
\begin{tikzpicture}
    \begin{axis}[
    scale only axis, 
    width=8cm,
    height=6.5cm,
    legend style={draw=none, at={(0.15,0.3)}, legend cell align={left}, fill=none, anchor=north},
	xmin=0e-6,
	xmax=1.5e-4,
	xtick distance=2.5e-5,
	ytick distance=0.1,
	ymin=-0.5,
	ymax=0.5,
	ytick={-0.5,-0.4,-0.3,-0.2,-0.1,0,0.1,0.2,0.3,0.4,0.5},
	yticklabels={-0.5,,,,,0,,,,,0.5},
	xtick={0,2.5e-5,5e-5,7.5e-5,1e-4,1.25e-4,1.5e-4},
 scaled x ticks=false,
	xticklabels={0,,,,,,150},
	xlabel={Time (\SI{}{\micro\second})},
	ylabel={Normalized pressure}, mark repeat={15},
	]
    

    \addplot+ [color=blue,smooth] file{figs/pressure_100kHz_receiver.dat};
    \addplot+ [color=teal,dashed] file{figs/pressure_200kHz_receiver.dat};
    \addplot+ [color=red,dotted] file{figs/pressure_500kHz_receiver.dat};
    \legend{\SI{100}{\kilo\hertz}, \SI{200}{\kilo\hertz}, \SI{500}{\kilo\hertz}}
    \end{axis}
    \end{tikzpicture}
    \caption{Normalized pressure response as a function of time at the receiver location of the fluid domain for \SI{100}{\kilo\hertz} (solid blue), \SI{200}{\kilo\hertz} (dashed teal), and \SI{500}{\kilo\hertz} (dotted red) frequencies.}
    \label{fig:pressure_receiver}
\end{figure}

Figure~\ref{fig:pressure_receiver} shows the normalized pressure (receiver pressure divided by the input pressure) as a function of time measured at the center of the receiver for the three frequencies. 
Since the curved surface is a free boundary, some waves travel through that surface to the receiver. As this travel distance is much smaller than the actual wave path (through the domain), the pulse through the surface reaches much earlier than the bulk waves as can be distinguished in the figure. However, their amplitudes are much smaller than the actual signal and can be discarded. The figure also shows that with the increase in the frequency, the bulk signal's amplitude increases whereas its spread decreases. This is expected since a high-frequency signal has a high energy and low diffraction. Additionally, although the signals arrive at different times, the difference in the arrival time is very small compared to the total travel time and hence it can be neglected. This also implies that these waves in the fluid medium experience insignificant dispersion. 
We compare the 3D wave behavior with a corresponding 2D model to investigate the effect of the 3rd dimension on wave propagation in the fluid domain, as discussed next.
\subsection{Comparison between 3D and 2D FEA}
\begin{figure}[!ht]
    \centering
    \def\svgwidth{0.85\linewidth}
	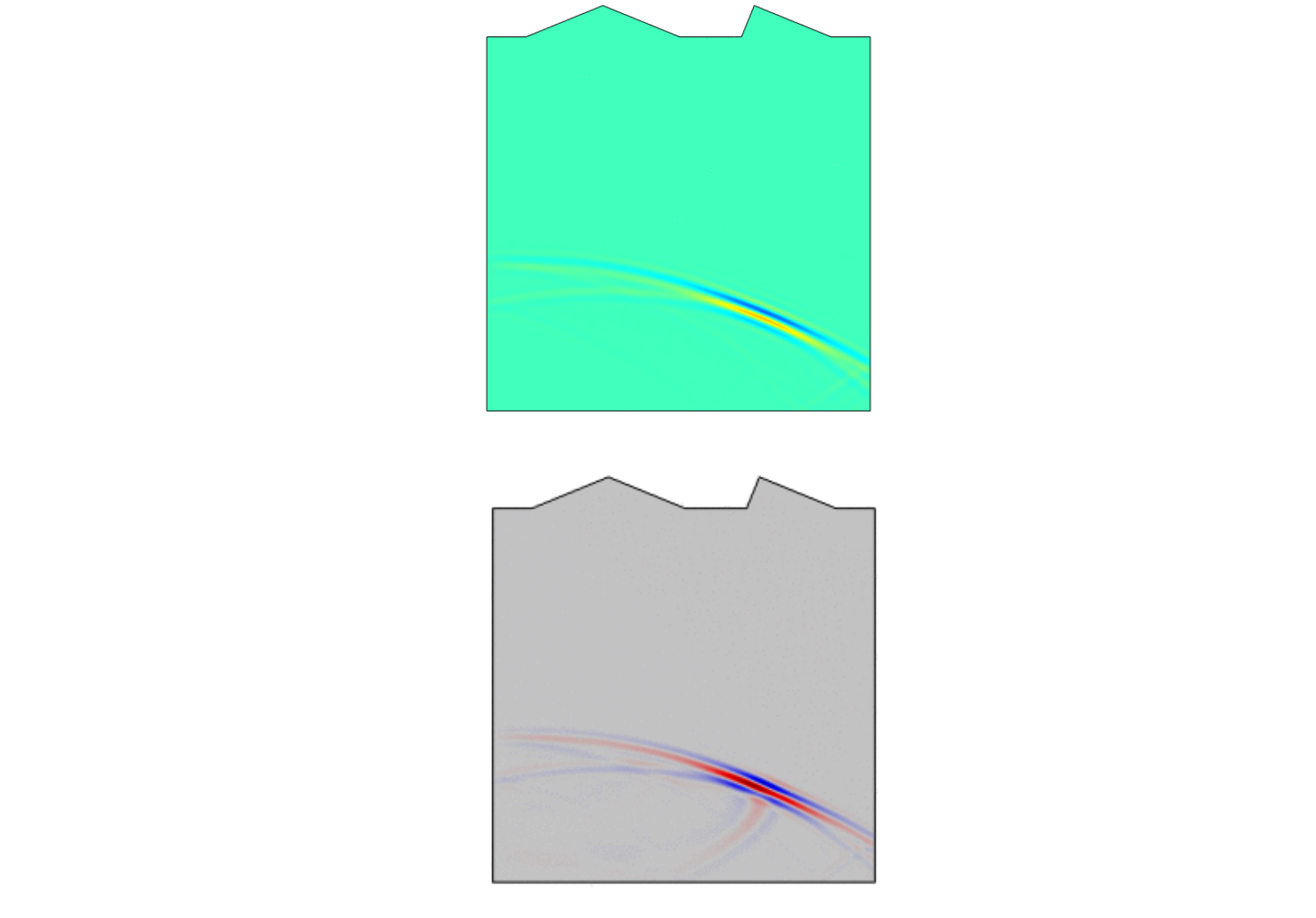
    \caption{Pressure profiles comparison between 2D and 3D at different instances at \SI{500}{\kilo\hertz}. The behavior is similar for 2D and 3D at the start of the wave propagation (a) and (d), and when the signal reaches the receiver (c), (f). However, as highlighted with black ellipses in (b) and (e), the wavefield during focusing is considerably different between 2D and 3D as the former cannot capture the focusing effect.}
    \label{fig:pressure_profiles_2D_500kHz}
\end{figure}

The 2D FEA model is built based on the geometry from Figure~\ref{fig:3D_model}, where only its central plane (with height $D$ and length $L_p$) is used. The incoming pressure is from a line source at an angle $\theta_3$ (similar to the plane source from Figure~\ref{fig:3D_model}), while the same time step (T/10) and total time (\SI{125}{\micro\second}) as in the 3D model are used here for consistency. We conducted the simulation for \SI{100}{\kilo\hertz}, \SI{200}{\kilo\hertz}, \SI{500}{\kilo\hertz}, and \SI{1}{\mega\hertz}. Figure~\ref{fig:pressure_profiles_2D_500kHz} compares the pressure profile between the 2D and the 3D analysis at \SI{500}{\kilo\hertz} for the same three instances discussed in Section~\ref{subsec:pressure_comparison_3D}.
From Figures~\ref{fig:pressure_profiles_2D_500kHz}(a) and~\ref{fig:pressure_profiles_2D_500kHz}(d), we can see that the wavefronts are very similar between 2D and 3D at the beginning of the simulation. However, the signal traveling back to the medium after impinging at the opposite side still shows a uniform wavefront in 2D, whereas in 3D we observe a focusing effect. This difference is highlighted in Figures~\ref{fig:pressure_profiles_2D_500kHz}(b) and \ref{fig:pressure_profiles_2D_500kHz}(d) using a black ellipse. The signals arriving at the receiver as shown in Figures~\ref{fig:pressure_profiles_2D_500kHz}(c) and \ref{fig:pressure_profiles_2D_500kHz}(f) show similar shapes meaning the focusing had little effect on the receiving signal's wavefront shape. To quantify the focusing and spreading effects on the signal amplitude, we compare the time response at the receiver between the 2D and 3D models.
\subsection{Pressure at the receiver comparison between 2D and 3D}
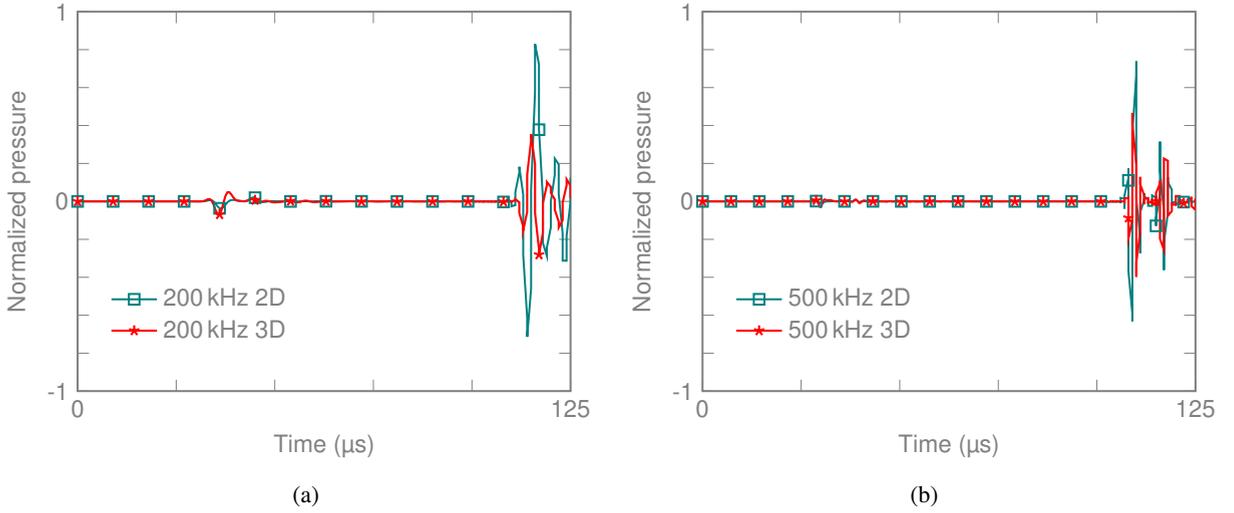
\begin{figure}[!ht]
 \begin{subfigure}[t]{0.49\textwidth}
      	\pgfplotsset{cycle list/Dark2}
	\def\svgwidth{1\linewidth}
  \pgfplotsset{cycle list/Dark2}
\begin{tikzpicture}
    \begin{axis}[
    scale only axis, 
    width=6.5cm,
    height=5cm,
    legend style={draw=none, at={(0.25,0.3)}, legend cell align={left}, fill=none, anchor=north},
	xmin=0e-6,
	xmax=1.25e-4,
	xtick distance=2.5e-5,
	ytick distance=0.1,
	ymin=-1,
	ymax=1,
	ytick={-1,-0.8,-0.6,-0.4,-0.2,0,0.2,0.4,0.6,0.8,1},
	yticklabels={-1,,,,,0,,,,,1},
	xtick={0,2.5e-5,5e-5,7.5e-5,1e-4,1.25e-4},
 scaled x ticks=false,
	xticklabels={0,,,,,125},
	xlabel={Time (\SI{}{\micro\second})},
	ylabel={Normalized pressure}, mark repeat={18},
	]
    

    \addplot+ [color=teal, mark=square] file{figs/pressure_200kHz_receiver_2D.dat};
    \addplot+ [color=red, mark=star, thick] file{figs/pressure_200kHz_receiver.dat};
    \legend{\SI{200}{\kilo\hertz} 2D, \SI{200}{\kilo\hertz} 3D}
    \end{axis}
    \end{tikzpicture}
	\subcaption{}
 \label{fig:pressure_200kHz}   
     \end{subfigure}
\begin{subfigure}[t]{0.49\textwidth}
	\centering
	\pgfplotsset{cycle list/Dark2}
	\def\svgwidth{0.5\linewidth}
  \pgfplotsset{cycle list/Dark2}
\begin{tikzpicture}
    \begin{axis}[
    scale only axis, 
    width=6.5cm,
    height=5cm,
    legend style={draw=none, at={(0.25,0.3)}, legend cell align={left}, fill=none, anchor=north},
	xmin=0e-6,
	xmax=1.25e-4,
	xtick distance=2.5e-5,
	ytick distance=0.1,
	ymin=-1,
	ymax=1,
	ytick={-1,-0.8,-0.6,-0.4,-0.2,0,0.2,0.4,0.6,0.8,1},
	yticklabels={-1,,,,,0,,,,,1},
	xtick={0,2.5e-5,5e-5,7.5e-5,1e-4,1.25e-4},
 scaled x ticks=false,
	xticklabels={0,,,,,125},
	xlabel={Time (\SI{}{\micro\second})},
	ylabel={Normalized pressure}, mark repeat={36},
	]
    

    \addplot+ [color=teal,mark=square] file{figs/pressure_500kHz_receiver_2D.dat};
    \addplot+ [color=red, mark=star, thick] file{figs/pressure_500kHz_receiver.dat};
    \legend{\SI{500}{\kilo\hertz} 2D, \SI{500}{\kilo\hertz} 3D}
    \end{axis}
    \end{tikzpicture}
	\subcaption{}
	\label{fig:pressure_500kHz}
\end{subfigure}%
\caption{Comparison of the normalized pressure at the receiver between 2D (teal line with square marker) and 3D (red line with asterisk marker) for (a) \SI{200}{\kilo\hertz} and (b) \SI{500}{\kilo\hertz} frequencies. The pulse shape and amplitudes are considerably different between 2D and 3D at \SI{200}{\kilo\hertz} while they tend to be closer at \SI{500}{\kilo\hertz}.}
\end{figure}

Normalized pressure at the receiver is calculated for the 2D models similar to the 3D case. Figure~\ref{fig:pressure_200kHz} and \ref{fig:pressure_500kHz} show the comparison of the normalized pressure between 2D and 3D for \SI{200}{\kilo\hertz} and \SI{500}{\kilo\hertz}, respectively. 
While the pulse shape at \SI{200}{\kilo\hertz} is considerably different between 2D and 3D, the response at \SI{500}{\kilo\hertz} is very close. This is also due to the lower diffraction at high frequencies. From this behavior, we can predict that at \SI{1}{\mega\hertz}, the variation between 2D and 3D will be even smaller and we will be able to use the 2D model to obtain the arrival time and the signal spread through the fluid medium with a nominal accuracy. The maximum amplitude of the signal is lower for the 3D model compared to the 2D because of the expansion in the third dimension. Additionally, we can see that this variation decreases with the increase in frequency as we discuss in detail next.
\subsection{Parametric response and dispersion relation}
\begin{figure}[!ht]
\begin{subfigure}[t]{0.49\textwidth}
      	\pgfplotsset{cycle list/Dark2}
    \centering
    \def\svgwidth{0.85\linewidth}
	\pgfplotsset{cycle list/Dark2}
\begin{tikzpicture}
    \begin{axis}[
    scale only axis, 
    width=6.5cm,
    height=5cm,
    legend style={draw=none, at={(0.75,0.3)}, legend cell align={left}, fill=none, anchor=north},
	xmin=0e-6,
	xmax=5000,
	xtick distance=1e3,
	ytick distance=1e-5,
	ymin=0,
	ymax=1e6,
	ytick={0,1e5,2e5,3e5,4e5,5e5,6e5,7e5,8e5,9e5,1e6},
	yticklabels={0,,,,,,,,,,1},
	xtick={0,1000,2000,3000,4000,5000},
 scaled x ticks=false,
 scaled y ticks=false,
	xticklabels={0,,,,,5000},
	xlabel={Wave number (\SI[per-mode=symbol]{}{1\per\meter})},
	ylabel={Frequency (\SI{}{\mega\hertz}}),
	]
    

    \addplot+ [color=teal,dashed, mark=star, style={fill=teal}] file{figs/dispersion_2D.dat};
    \addplot+ [color=red, mark=*,] file{figs/dispersion_3D.dat};
    \legend{2D, 3D}
    \end{axis}
    \end{tikzpicture}
    \subcaption{}
    \label{fig:dispersion_relation}
\end{subfigure}
\begin{subfigure}[t]{0.49\textwidth}
	\centering
	\pgfplotsset{cycle list/Dark2}
    \centering
    \def\svgwidth{0.85\linewidth}
	\pgfplotsset{cycle list/Dark2}
\begin{tikzpicture}
    \begin{axis}[
    scale only axis, 
    width=6.5cm,
    height=5cm,
    legend style={draw=none, at={(0.75,0.3)}, legend cell align={left}, fill=none, anchor=north},
	xmin=0e-6,
	xmax=1e6,
	xtick distance=2e5,
	ytick distance=0.2,
	ymin=0,
	ymax=1.2,
	ytick={0,0.2,0.4,0.6,0.8,1,1.2},
	yticklabels={0,,,,,,1.2},
	xtick={0,1e5,2e5,3e5,4e5,5e5,6e5,7e5,8e5,9e5,1e6},
 scaled x ticks=false,
 scaled y ticks=false,
	xticklabels={0,,,,,0.5,,,,,1},
	xlabel={Frequency (\SI{}{\mega\hertz})},
	ylabel={Pressure ratio},
	]
    \addplot[color=red,
   only marks,
   mark=triangle,
   ] coordinates {
    (1e6,1.01)
    };

    \addplot+ [color=teal, mark=star] file{figs/pressure_ratio.dat};
    \addplot+ [color=teal,dashed, mark=star] file{figs/pressure_ratio2.dat};
    \legend{Ray tracing, FEM}
    \end{axis}
    \end{tikzpicture}
    \subcaption{}
    \label{fig:pressure_ratio}
    \end{subfigure}
    \caption{(a) Dispersion relation (b) Pressure ratio comparison between 2D and 3D. The dispersion relations for 2D and 3D are linear (non-dispersive) and closely matching. The pressure ratio increases with frequency and at low frequencies the variation is higher due to high diffraction. The variation at high frequencies would be more linear and proportional to frequency. The extrapolated pressure ratio at \SI{1}{\mega\hertz} predicted by FEM (teal asterisk) is close to the ray acoustic estimation shown using a red triangle (\SI{3.6}{\percent} variation).}
\end{figure}
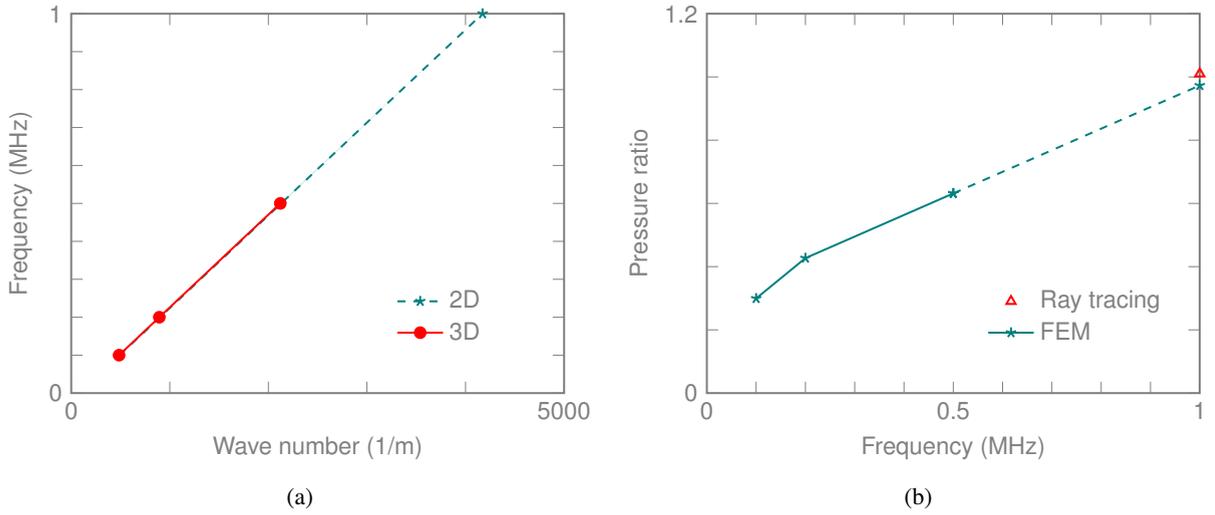

Similar to the ray tracing discussed in Section~\ref{subsec:ray_tracing}, we can obtain the scaling factor that determines the variation in pressure amplitude in the fluid domain due to 3D via FEA. To that end, we need to determine the dispersion relation of the domain in addition to the time response shown in Figures~\ref{fig:pressure_200kHz} and~\ref{fig:pressure_500kHz}. The dispersion relation relates the frequency with the wave vector (magnitude is the reciprocal of the wavelength and directed towards the wavefront), whose slope provides the wave speed~\cite{graff2012wave}. If the wave speed varies with the frequency, then the medium is dispersive. Since the domain for all these three frequency cases is the same, the travel distance remains constant. The arrival time is then obtained from the time response (Figure~\ref{fig:pressure_500kHz}), which  is used to calculate the average sound speed and the magnitude of the wave vector within the medium as follows
\begin{align}
    v_m=\frac{L_{tot}}{t_a}, && \lambda=\frac{v_m}{f}, && k=\frac{2\pi}{\lambda},
\label{eq:dispersion}
\end{align}
where, $l_{tot}$ is the total distance from the source to the receiver, $t_a$ is the arrival time of the pulse, $v_m$ is the calculated sound speed within the domain, $\lambda$ is the wavelength, and $k$ is the magnitude of the wave vector also called a wave number. Figure~\ref{fig:dispersion_relation} shows the comparison of the dispersion relation between 
2D and 3D FEA, which are linear and very similar. In other words, the fluid medium can be considered non-dispersive and the additional dimension does not have a significant influence on the dispersion relation of the medium. 

Figure~\ref{fig:pressure_ratio} shows the relation between the pressure ratio for different applied frequencies, which is calculated by dividing the peak pressure values of 3D with 2D obtained from the corresponding time response plots. We see that the pressure ratio increases with frequency and the slope between \SI{100}{\kilo\hertz} and \SI{200}{\kilo\hertz} is larger than the slope between \SI{200}{\kilo\hertz} and \SI{500}{\kilo\hertz} because at lower frequencies due to higher diffraction, the signal experiences larger spread and thus the losses increases. At higher frequencies, the behavior would be more linear due to the lower loss. Additionally, the sound pressure within the fluid is directly related to the sound speed and frequency, where the former can be considered constant due to the non-dispersive behavior. Thus, we can linearly extrapolate the pressure ratio in Figure~\ref{fig:pressure_ratio} from \SI{500}{\kilo\hertz} to \SI{1}{\mega\hertz}, and its value is 0.973. Recall that we estimated the pressure ratio from ray tracing to be 1.01 (marked using a red triangle in the figure), which is only \SI{3.6}{\percent} higher than the FEA prediction. We proceed to determine the scaling factor for the pipe wall using the geometric projection method, as discussed next.
\section{Geometric projection of the pipe wall}
\begin{figure}[!ht]
    \centering
    \def\svgwidth{0.85\linewidth}
	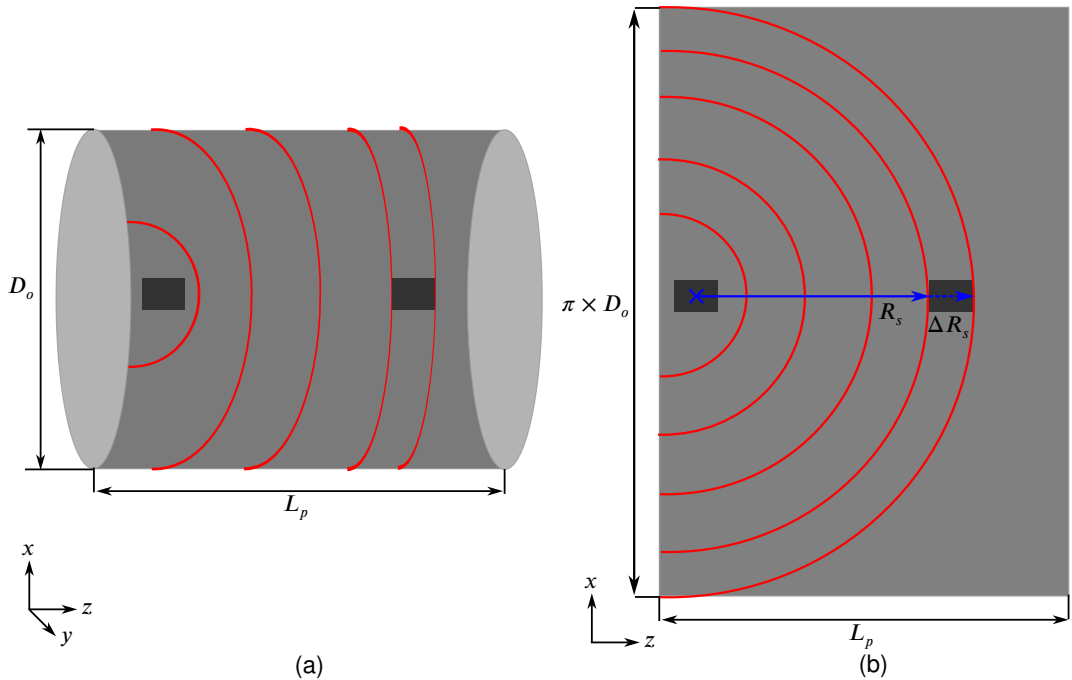
    \caption{Geometric projection of waves traveling in the pipe wall (a) top view of the pipe wall with wedges where $D_o$ and $L_p$ are marked. It shows cylindrical waves traveling from the transmitter to the receiver. (b) projection of the pipe to a plane where its width is the same as the length of the pipe ($L_p$), while its height is the circumference of the pipe ($\pi\times D_o$). The wavefield is represented using concentric circles and the blue arrow with length $R_s$ shows the radius of the field when the waves impinge the receiver. $\Delta R_s$ is the change in the radius of the wavefield from the start to the end of the receiver that is equivalent to $b_w$.}
    \label{fig:geometric_projection}
\end{figure}

Geometric projection is carried out by unwrapping the cylindrical pipe wall to a plane and measuring the area of the expanding semi-circle when reaching the receiver as shown in Figure~\ref{fig:geometric_projection}. 
The pipe with the outer diameter $D_o$ and length $L_p$ is projected to a rectangle with width $L_p$ and height $\pi\times D_o$ as shown in the figure. The expanding cylindrical wavefield can be represented as concentric circles in the planar domain, which is shown using red semi-circular rings in Figure~\ref{fig:geometric_projection}(b). The blue arrow of length $R_s$ is the radius of the expanding circle when reaching the receiver and $\Delta R_s$ is its variation from the start to the end of the receiver (along $z$ direction).

On the pipe wall, solid waves are present, whose energy is directly related to the area of the circle. We use the area of the receiver and the total area of the circular strip enclosing the receiver to obtain the area ratio as follows:
\begin{align}
    A_w=b_w\times t_w, && A_s=\pi\times \left[\left(R_s+\Delta R_s\right)^2-R_s^2\right], && A_r=A_w/A_s,
    \label{eq:geometric_ratio}
\end{align}
where $A_w$ is the area of the receiver, $A_s$ is the area of the strip encircling the receiver, and $A_r$ is the area ratio, which is equivalent to the energy ratio. For the current geometry, $R_s=\SI{70}{\milli\meter}$ and $\Delta R_s=\SI{36}{\milli\meter}$ resulting in $A_r=0.05426$. As the acoustic energy is related to the square of the displacement (in the solid domain), the displacement ratio $d_r=\sqrt{A_r}=0.233$ on the pipe wall. In other words, all waves traveling on the pipe wall from the transmitter to the receiver (with cylindrically expanding wavefield), experience a reduction in amplitude with the factor $d_r$ when the effect of the 3\textsuperscript{rd} dimension is incorporated into the 2D analysis.
We now apply scaling factors in both fluid and solid to the output signal from the clamp-on flowmeter. To that end, we use the complete 2D clamp-on model discussed in Section 3.3 of Chapter 6, as follows. 
\section{2D wave propagation of the clamp-on flowmeter}
\begin{figure}[!ht]
    \centering
    \def\svgwidth{1\linewidth}
	\input{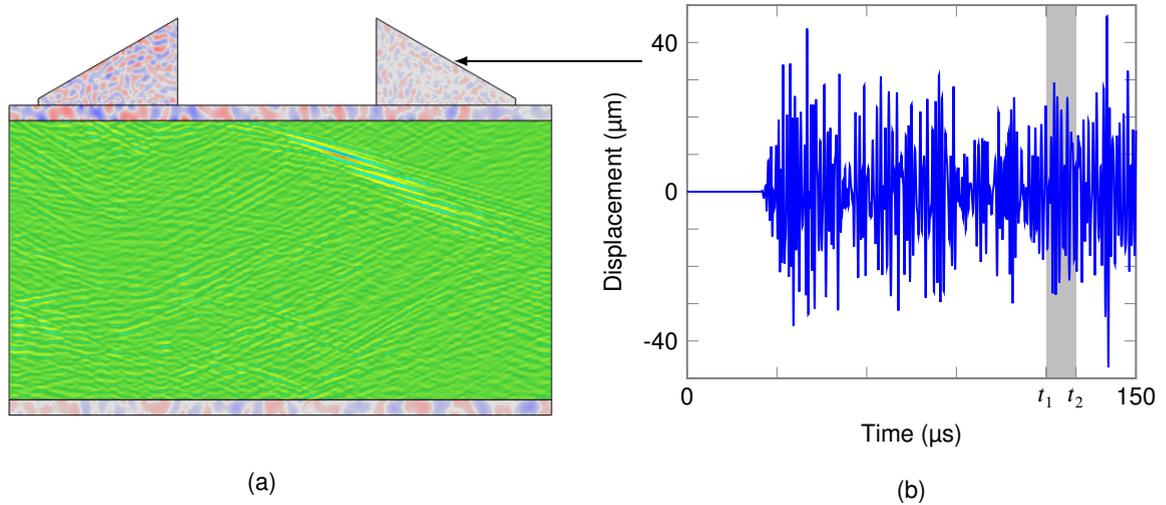}
    \caption{2D clamp-on flowmeter response (a) snapshot of the pressure profile when the fluid signal is about to impinge the pipe wall (b) the displacement (in $\SI{}{\micro\meter}$) at the receiver as a function of time (in $\SI{}{\micro\second}$). The gray shaded region corresponds to the arrival time of the fluid signal bounded by $t_1=\SI{120}{\micro\second}$ and $t_2=\SI{130}{\micro\second}$.}
    \label{fig:2d_clamp_on_system}
\end{figure}

The 2D time-domain analysis of the complete clamp-on flowmeter is performed using dG FEM and the results are provided in Figure~\ref{fig:2d_clamp_on_system}. 
Figure~\ref{fig:2d_clamp_on_system}(a) shows the pressure profile of the clamp-on flowmeter when the signal from the fluid domain is about to impinge the pipe wall. Here, we can see that the wedge, pipe wall, and fluid are very noisy, as can be seen by the output displacement measured on the receiver shown in Figure~\ref{fig:2d_clamp_on_system}(b). The arrival time of the working signal bounded between $t_1=\SI{120}{\micro\second}$ and $t_2=\SI{130}{\micro\second}$ is determined by using the sound speeds and distance between the wedges and is highlighted using the gray-shaded region in the same figure. The noise in the output signal is majorly contributed by the multiple ringing effects from the transmitting wedge, i.e., the input pulse after reflecting from the wedge-pipe wall interface travels back to the wedge. This pulse then reflects from other free surfaces (of the wedge) back to the pipe wall and travels through the wall as noise signals. This process repeats until the amplitude diminishes to the device's noise floor. Another noise source is the ringing of waves between the transmitter and the receiver. Since the distance between the wedges is fixed (\SI{52}{\milli\meter}), the time difference between these ringing can be easily calculated using the sound speed in the pipe wall and the travel distance (twice the distance between the wedges). The ringing interval between transmitter and receiver is obtained as \SI{21.1}{\micro\second} for P waves whereas \SI{33.5}{\micro\second} for S waves. However, as these noise signals are present throughout the recorded time it would be difficult to separate signals. To that end, we perform the time-domain analysis of the clamp-on flowmeter with certain modifications as discussed next.
\subsection{Output signal segmentation of the clamp-on ultrasonic flowmeter}
\begin{figure}[!ht]
    \centering
    \def\svgwidth{1\linewidth}
	\input{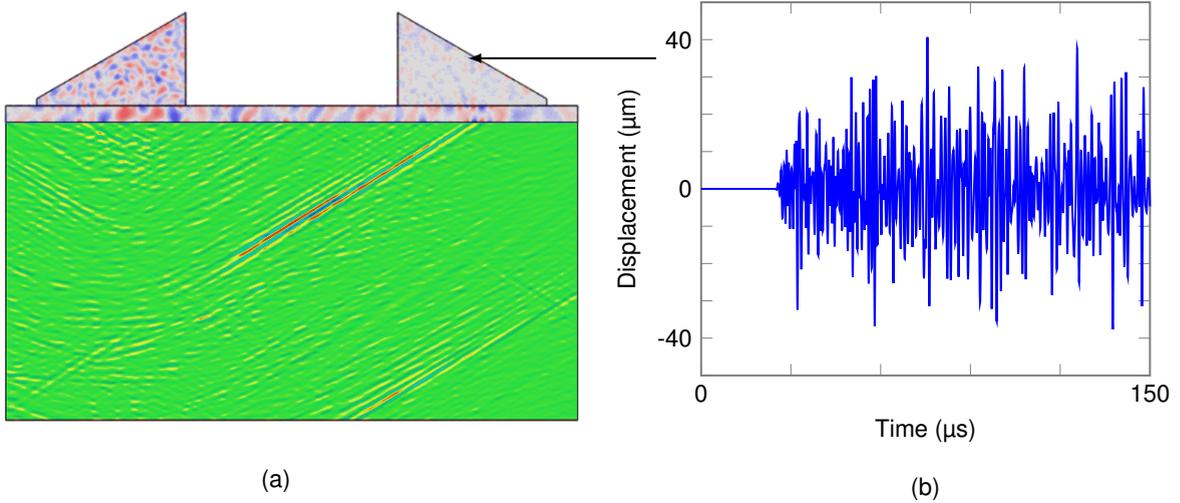}
    \caption{Response of the clamp-on flowmeter model without the bottom pipe wall (a) snapshot of the pressure profile for the same time instance as Figure~\ref{fig:2d_clamp_on_system}(a), (b) the receiving wedge's displacement as a function of time also measured at the same location as Figure~\ref{fig:2d_clamp_on_system}(b).}
    \label{fig:clamp_on_solid}
\end{figure}
\begin{figure}[!ht]
\begin{subfigure}[t]{0.49\textwidth}
      	\pgfplotsset{cycle list/Dark2}
    \centering
    \def\svgwidth{0.85\linewidth}
	  
\begin{tikzpicture}
	\begin{axis}[height=6.5cm, 
							 width=7.5cm, 
							 x tick label style={text=black},
    x label style={text=black},
    y tick label style={text=black},
    y label style={text=black},
    title style ={text=black},
							xmin=0,
	xmax=1.5e-4,
	xtick distance=3e-5,
	ytick distance=0.1,
	ytick distance=1e-5,
	ymin=-2e-5,
	ymax=2e-5,
	ytick={-2e-5,-1.5e-5,-1e-5,-5e-6,0,5e-6,1e-5,1.5e-5,2e-5},
    yticklabels={-20,,,,0,,,,20},
		scaled x ticks=false,
	xtick={0,3E-5,6E-5,9E-5,1.2E-4,1.5E-4},
	xticklabels={0,,,,,150},
	xlabel={Time (\SI{}{\micro\second})},
	ylabel={Displacement (\SI{}{\micro\meter})},
 extra x ticks = {1.2e-4,1.32e-4},
	extra x tick labels = {$t_1$,$t_2$},
	title style={at={(0.5,-0.4)},anchor=south},
	]
\draw [gray, fill=gray, opacity= 0.5, draw opacity =0] (axis cs:1.2e-4,-4e-5)  rectangle (axis cs:1.3e-4,4e-5);
    \addplot+ [color=blue, mark=none] file{figs/Time_response_fluid_signal.dat}; 

				\end{axis} 
 
\end{tikzpicture}
    \subcaption{}
    \label{fig:fluid_signal}
\end{subfigure}
\begin{subfigure}[t]{0.49\textwidth}
	\centering
	\pgfplotsset{cycle list/Dark2}
    \centering
    \def\svgwidth{0.85\linewidth}
	  
\begin{tikzpicture}
	\begin{axis}[height=6.5cm, 
							 width=7.5cm, 
							 x tick label style={text=black},
    x label style={text=black},
    y tick label style={text=black},
    y label style={text=black},
    title style ={text=black},
							xmin=0,
	xmax=1.5e-4,
	xtick distance=3e-5,
	ytick distance=0.1,
	ytick distance=1e-5,
	ymin=-2e-5,
	ymax=2e-5,
	ytick={-2e-5,-1.5e-5,-1e-5,-5e-6,0,5e-6,1e-5,1.5e-5,2e-5},
    yticklabels={-20,,,,0,,,,20},
		scaled x ticks=false,
	xtick={0,3E-5,6E-5,9E-5,1.2E-4,1.5E-4},
	xticklabels={0,,,,,150},
	xlabel={Time (\SI{}{\micro\second})},
	ylabel={Displacement (\SI{}{\micro\meter})},
	title style={at={(0.5,-0.4)},anchor=south},
	]
\draw [gray, fill=gray, opacity= 0.5, draw opacity =0] (axis cs:1.2e-4,-4e-5)  rectangle (axis cs:1.3e-4,4e-5);
    \addplot+ [color=blue, mark=none] file{figs/Time_response_scaled_signal.dat}; 
				\end{axis} 
 
\end{tikzpicture}
    \subcaption{}
    \label{fig:scaled_response}
    \end{subfigure}
    \caption{(a) Displacement response of the signal traveling through the fluid domain calculated by subtracting the signal from the solid domain (Figure~\ref{fig:clamp_on_solid}(b)) from the time response of the clamp-on flowmeter (Figure~\ref{fig:2d_clamp_on_system}(b)), where the shaded region represents the working signal bounded by $t_1$ and $t_2$. (b) The displacement at the receiver was obtained after scaling the solid domain's signal by $d_r=0.233$, and the fluid domain's signal by $p_r=1.01$ and combining both. The gray-shaded region is the same as (a).}
\end{figure}
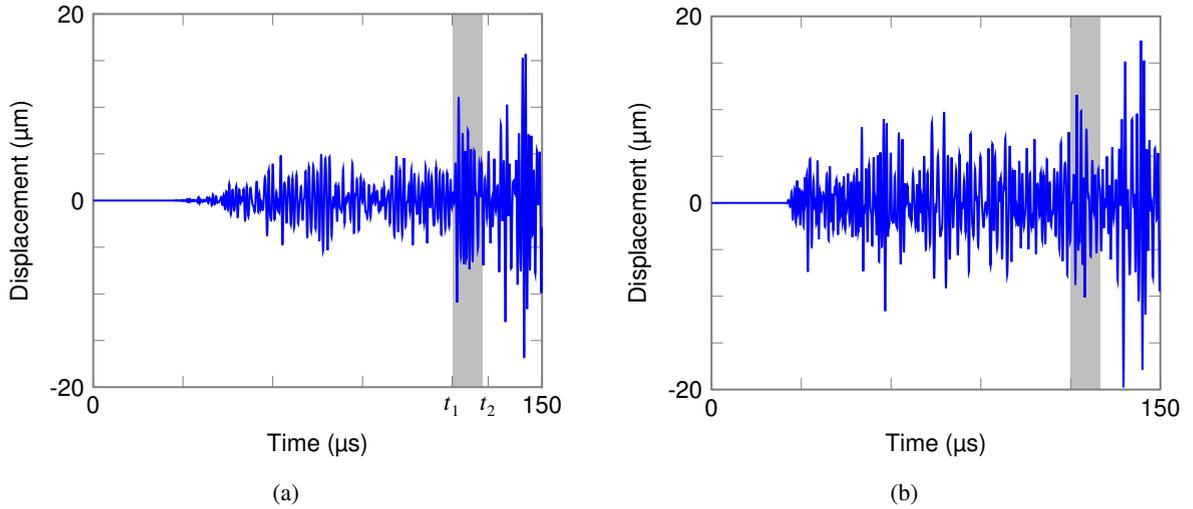

The geometry of the 2D FEM model is changed such that the bottom pipe wall is replaced by an impedance BC to ensure that the required signal from the fluid domain does not reach the receiver. The pressure profile shown in Figure~\ref{fig:clamp_on_solid}(a) is at the same instance as in Figure~\ref{fig:2d_clamp_on_system}(a) where we can clearly see that no signal arrives at the receiver from the fluid. The corresponding time response is also extracted at the top of the wedge (same location as the response from Figure~\ref{fig:2d_clamp_on_system}(b)) and plotted in Figure~\ref{fig:clamp_on_solid}(b). Since we assume linearity throughout the modeling and we kept all other factors (geometry, simulation settings, and solver settings) unchanged, we can safely assume that the difference between the signal from the clamp-on model and the model without the bottom pipe wall will provide the required fluid signal as shown in Figure~\ref{fig:fluid_signal}. Here, we can see that when the influences of the solid portion and the solid-fluid interaction (of the top pipe wall-fluid interface) are removed, the signal from the fluid is clearly visible and as predicted it is between \SI{120}{\micro\second} and \SI{130}{\micro\second}. Additionally, the signal amplitude is also diminished by a factor of two due to the lack of waves from the solid domain. This also implies that the waves traveling in the solid domain possess higher energy than the required signal from the fluid. We can also observe an additional signal just after the required signal, which is due to the secondary reflection from the transmitting wedge. 

We apply the different scaling factors derived in previous sections to these two (noise and required signals) and add them together to obtain the complete output response of the clamp-on ultrasonic flowmeter. To that end, the signal from the fluid is scaled with the pressure ratio $p_r=1.01$ and the signal from the solid region is scaled with the displacement ratio $d_r=0.233$. The assumption here is that, since we already consider the area of the receiving wedge while determining $p_r$ (from ray acoustics), the scaling factor stays the same when the fluid signal travels through the pipe wall and then to the wedge because of the short travel time. In other words, the displacement scaling factor for the fluid signal is equal to $p_r$. The resulting displacement response of the receiving wedge is shown in Figure~\ref{fig:scaled_response}. Although the working signal can be identified, the signals from the solid portion and multiple solid-fluid interactions are still present. These are not spurious signals since due to the multiple solid-fluid interfaces these signals can be generated in a clamp-on ultrasonic flowmeter. However, in most commercially used clamp-on ultrasonic flowmeters, plastic wedges are used to transmit and receive signals. They possess damping due to viscoelastic behavior that enables them to minimize the secondary reflections and ringing between the wedges. Incorporating this complex behavior needs the complete viscoelastic characterization of the wedge materials, which is out of the scope of this study. However, compared to Figure~\ref{fig:2d_clamp_on_system}(b), the response of the clamp-on ultrasonic flowmeter is better captured here due to incorporating the influence of 3D via different scaling factors. 
\section{Summary and Conclusion}
To characterize the wave propagation through a complex media such as a clamp-on ultrasonic flowmeter, we proposed a semi-analytical approach combining 2D and 3D FEA, ray tracing, and geometric projection. Since the complete 3D transient analysis of the clamp-on flowmeter at high frequency is computationally extremely intensive and as we cannot rely on frequency-domain analyses because they do not represent the physical reality, we analyzed the clamp-on system using 2D transient FEA. We then separated the geometry into different domains, where we characterized the fluid domain with 3D FEA and ray tracing, and the solid domain via the geometric projection. We determined scaling factors that account for the effects of 3D for waves traveling in different domains. Using these scaling factors, we scale the output response at the receiver from the 2D analysis allowing us to distinguish the fluid signal from the surrounding noise, which was impossible otherwise. We conclude that:
\begin{itemize}
    \item To represent the wave propagation through a clamp-on flowmeter accurately, it is beneficial to separate the domain into different sub-domains and analyze them individually. This aids in reducing the overall computational cost and allows us to section the output signal into different parts;
    \item Dispersion relation and the wave envelope shape are comparable between 2D and 3D FEA at low frequencies implying that low-frequency models could provide reasonable predictions and allow us to forecast the wave propagation behavior of the fluid domain at high frequencies by extrapolation;
    \item Ray tracing assumes a particle-like behavior and hence can provide accurate descriptions at high frequencies. For an isotropic solid domain without dispersion and damping, depending on the type of wavefield, geometrical projection based on analytical expressions can be used to characterize the wave propagation behavior. We can use these two approaches in conjunction with 2D FEA to represent the wave propagation through clamp-on flowmeters accurately;
\end{itemize}
The proposed method lacks the means to model the solid-fluid interface in 3D, so we cannot properly characterize the interfacial waves; however, with the 2D analysis, we can partially get the interface behavior. Since in the selected clamp-on flowmeter, the influence of the interface waves was not as prominent as the solid and fluid waves, we can ignore them. Additionally, the viscoelastic damping behavior of the plastic wedges is also not included in the proposed model, which influences its accuracy. Thus, the next step could be to characterize the viscoelastic behavior of the wedge materials and include them in the wave propagation model. Another possible direction could be to accurately model the solid-fluid interface and generalize the approach for more complex systems' wave propagation behavior at high frequencies.   
\printcredits


\bibliography{cas-refs}
\end{document}